\documentstyle[12pt]{article}

\def\JHEP #1 #2 #3{{\sl Journ. High. Energy. Phys.} {\bf#1} (#2) #3}

\def\PRL #1 #2 #3{{\sl Phys. Rev. Lett.} {\bf#1} (#2) #3}

\def\NPB #1 #2 #3{{\sl Nucl. Phys.} {\bf B#1} (#2) #3}

\def\NPBP #1 #2 #3{{\sl Nucl. Phys. B (Proc. Suppl.) } {\bf B#1} (#2) #3}

\def\NPBFS #1 #2 #3 #4{{\sl Nucl. Phys.} {\bf B#2} [FS#1] (#3) #4}

\def\CMP #1 #2 #3{{\sl Commun. Math. Phys.} {\bf #1} (#2) #3}

\def\PRD #1 #2 #3{{\sl Phys. Rev.} {\bf D#1} (#2) #3}

\def\PLA #1 #2 #3{{\sl Phys. Lett.} {\bf #1A} (#2) #3}

\def\PLB #1 #2 #3{{\sl Phys. Lett.} {\bf #1B} (#2) #3}

\def\JMP #1 #2 #3{{\sl Journ. Math. Phys.} {\bf #1} (#2) #3}

\def\PTP #1 #2 #3{{\sl Prog. Theor. Phys.} {\bf #1} (#2) #3}

\def\SPTP #1 #2 #3{{\sl Suppl. Prog. Theor. Phys.} {\bf #1} (#2) #3}

\def\AoP #1 #2 #3{{\sl Ann. of Phys.} {\bf #1} (#2) #3}

\def\PNAS #1 #2 #3{{\sl Proc. Natl. Acad. Sci. USA} {\bf #1} (#2) #3}

\def\RMP #1 #2 #3{{\sl Rev. Mod. Phys.} {\bf #1} (#2) #3}

\def\PR #1 #2 #3{{\sl Phys. Reports} {\bf #1} (#2) #3}

\def\AoM #1 #2 #3{{\sl Ann. of Math.} {\bf #1} (#2) #3}

\def\UMN #1 #2 #3{{\sl Usp. Mat. Nauk} {\bf #1} (#2) #3}

\def\FAP #1 #2 #3{{\sl Funkt. Anal. Prilozheniya} {\bf #1} (#2) #3}

\def\FAaIA #1 #2 #3{{\sl Functional Analysis and Its Application} {\bf

#1} (#2) #3}

\def\BAMS #1 #2 #3{{\sl Bull. Am. Math. Soc.} {\bf #1} (#2)

#3} \def\TAMS #1 #2 #3{{\sl Trans. Am. Math. Soc.} {\bf #1} (#2) #3}

\def\InvM #1 #2 #3{{\sl Invent. Math.} {\bf #1} (#2) #3}

\def\LMP #1 #2 #3{{\sl Letters in Math. Phys.} {\bf #1} (#2) #3}

\def\IJMPA #1 #2 #3{{\sl Int. J. Mod. Phys.} {\bf A#1} (#2) #3}

\def\AdM #1 #2 #3{{\sl Advances in Math.} {\bf #1} (#2) #3}

\def\RMaP #1 #2 #3{{\sl Reports on Math. Phys.} {\bf #1} (#2) #3}

\def\IJM #1 #2 #3{{\sl Ill. J. Math.} {\bf #1} (#2) #3}

\def\CMP #1 #2 #3{{\sl Ill. Commun. Math. Phys.} {\bf #1} (#2) #3}

\def\APP #1 #2 #3{{\sl Acta Phys. Polon.} {\bf #1} (#2) #3}

\def\TMP #1 #2 #3{{\sl Theor. Mat. Phys.} {\bf #1} (#2) #3}

\def\JPA #1 #2 #3{{\sl J. Physics} {\bf A#1} (#2) #3}

\def\JSM #1 #2 #3{{\sl J. Soviet Math.} {\bf #1} (#2) #3}

\def\MPLA #1 #2 #3{{\sl Mod. Phys. Lett.} {\bf A#1} (#2) #3}

\def\JETP #1 #2 #3{{\sl Sov. Phys. JETP} {\bf #1} (#2) #3}

\def\JETPL #1 #2 #3{{\sl  Sov. Phys. JETP Lett.} {\bf #1} (#2) #3}

\def\PHSA #1 #2 #3{{\sl Physica} {\bf A#1} (#2) #3}

\def\CQG #1 #2 #3{{\sl Class. Quantum Grav.} {\bf #1} (#2) #3}

\def\SJNP #1 #2 #3{{\sl Sov. J. Nucl. Phys. (Yadern.Fiz.)} {\bf #1} (#2) #3}
\def\FdP #1 #2 #3{{\sl Fortschr. Phys.} {\bf A#1} (#2) #3}
\def\SSC #1 #2 #3{{\sl Solid  State Commun.} {\bf #1} (#2) #3}

\def\a{\alpha}
\def\b{\beta}
\def\g{\gamma}
\def\d{\delta}
\def\e{\varepsilon}
\def\r{\varrho}
\def\k{\kappa}
\def\l{\lambda}
\def\L{\Lambda}
\def\s{\sigma}
\def\S{\Sigma}
\def\Th{\Theta}
\def\th{\theta}

\def\Om{\Omega}

\def\t{\tau}

\def\g{\gamma}
\newcommand{\f}[1] {(\ref{#1})}

\begin{document}
\thispagestyle{empty}

\medskip
\begin{center}
{\large\bf Hamiltonian of Tensionless Strings with Tensor Central Charge Coordinates}\\
\vspace{10mm}
 A.A. Zheltukhin$^{\rm a,b}$ and U. Lindstr\"om$^{\rm a} $
\end{center}
\begin{center}
$^{a}$ Institute of Theoretical Physics, University of Stockholm , SCFAB,\\
 SE-106 91 Stockholm, Sweden \\
$^{b}$ Kharkov Institute of Physics and Technology, 61108, Kharkov,  Ukraine
\vskip 2.cm

{\bf Abstract}
\end{center}

\vspace{10mm}

 A new class of twistor-like string models in four-dimensional space-time extended by the addition of six tensorial central charge (TCC) coordinates $z_{mn}$ is studied.

The Hamiltonian of tensionless string in the extended space-time is derived and its symmetries are investigated. We establish that the string constraints reduce the number of independent TCC coordinates $z_{mn}$ to one real effective coordinate which composes an effective 5-dimensional target space together with the $x^{m}$ coordinates.
We construct the P.B. algebra of the first class constraints and discover that  it coincides with the P.B. algebra of tensionless strings.
 The Lorentz covariant antisymmetric Dirac $\hat{\mathbf C}$-matrix of the P.B. of the second class constraints is constructed and its algebraic structure is further presented.

\vspace{10mm}

\section {Introduction}

The supersymmetry algebra with tensor central charges \cite{dafre,holpro,ziz}
  is relevant for the brane sector \cite{hupol,azto,dust} of M/string theory \cite{wit,pol1,town2}. It is consistent with an extension of space time by 
 tensorial central charge (TCC)  \cite{curt,gren,hew,eis}. A wide of relevant models were constructed in \cite{sieg,bes,se} where the coordinates $z_{m_1m_2...m_p}$ are considered as new independent degrees of freedom corresponding to the p-form TCC generators $Z_{m_1m_2...m_p}$.
A  central charge carried by the BPS brane/string preserving $1/2$ of the N=1 supersymmety \cite{azgit} appears in QCD \cite{ds,gs} due to spontaneous breakdown of discrete chiral symmetry \cite{vy} and it is associated with a  domain wall created by the gluino condensate. The TCC extension of the superparticle \cite{ruds,balu} and superbrane \cite{ggt,gah} models has led to new solutions which
define the fraction  of spontaneously broken supersymmetries.
In \cite{gght} combinations of momentum and domain-wall charges were used to 
characterize the BPS state spectrum of preserved fractions of $D=4$ $ N=1$ supersymmetry. The interesting physical and mathematical foundations for the central extension of  superspace were considered in \cite{caib}, where  a connection  between the topological charges and TCC coordinates was analysed. 

All the above points to the relevance of studying the dynamical role of the TCC coordinates in string/brane models. To this end new models for strings moving in $D=4$ space-time extended by six real coordinates  $z_{mn}$  corresponding to the TCC charges $Z_{mn}$ were constructed in \cite{zli}. The action integrals in these models generalize a twistor like formulations  \cite{gz} for the Nambu-Goto and tensionless string actions. The suggested models have a natural supersymmetric generalization and may be considered  as a bosonic sector of tensile or tensionless superstrings moving in the extended space-time.                          Keeping in mind the mechanisms for the generation of tension \cite{zi,town1,belt,haliu,gz} induced by the interaction  of tensionless strings \cite{sch,kl,z1}, branes \cite{z2} or D-branes \cite{lir} with additional coordinates or fields one could expect a similar effect in the presense of the TCC coordinates. 

Therefore, in  \cite{zli} the action  of tensionless string minimally extended by the introduction of the term linear in the derivatives of TCC coordinates was choosen for study.
It was shown that inclusion of the $z^{mn}$ coordinates lifts  the light-like character of the tensionless string worldsheet and removes the degeneracy of the worldsheet metric. This could be treated as a hint of string tension being generated. 
A  particular set of solutions for the system of the string equations and integrability conditions found in \cite{zli} describes the string motion free of transverse oscillations in the $x-$directions  and a wave process in the  $z-$directions. These particular solutions do not capture full
effects due to the TCC coordinates.

However, these effects have to be visible at the level of the generalized Virasoro algebra for the local symmetry generators of the model.
 To derive this algebra we need to construct the Hamiltonian formalism and that is the objective of the present paper.  

By solving this problem the Hamiltonian and the constraints of the tensionless string in the extended space-time are constructed. After a covariant separation of the constraints into first and second classes we find that only $10$ phase space variables of  $36$ are independent. This corresponds to the string 
moving in an effective  $(4+1)-$dimensional target space instead of the  primary $(4+6)-$dimensional space-time. As described in section 5 we expect this effective target space to be $AdS_{5}$.
The P.B. algebra for the first class constraints is isomorphic 
to the corresponding P.B. algebra for tensionless string. This algebra has a structure similar to that of the contracted algebra of rotations of (Anti) de Sitter space. However, the second  class constraints in the model under question differ from the corresponding constraints for the tensionless string and therefore 
encode the physical effects of the TCC coordinate. 
The second class constraints will deform the original P.B. algebra of the first class constraint into the Dirac bracket algebra and restore the correct equations of motion. Then the D.B. algebra of  the first class constraints  together with the string  equations will define the structure of the string world sheet and the effective 5-dimensional target-space.
 (Equivalently, one can use a formalism where the second class constraints 
are converted into first class constraints in an extended phase space \cite{fs,bff,em}.)

We construct the Lorentz covariant antisymmetric Dirac $\hat{\mathbf C}$-matrix of the P.B. of the second class constraints displaying its algebraic structure.
 It may be presented in a condenced form as a 9x9 complex matrix. 

 \section {Strings with tensor central charge coordinates}

To describe the string dynamics we start from a twistor-like representation of the tensile/tensionless string action \cite{gz}
\begin{equation}\label{1}
S=\k\int (p_{mn}\,d{x^m}{\wedge}\,d{x^n}+ \L),
\end{equation}
where the local bivector $ p_{mn}(\t,\s)$  is composed of the local Newman-Penrose dyads attached to the worldsheet and the $\L-$term  fixes the orthonormality constraint  for the spinorial dyads (or twistor like variables). 

For the case of tensionless string $ p_{mn}(\t,\s)$ should be a null bivector defined by the condition 
\begin{equation}\label{2}
p_{mn}p^{mn}=0\quad,  \qquad
{\eta_{mn}=(-+++)},
\end{equation}
which implies the general solution for $ p_{mn}(\t,\s)$ in the form of a bilinear covariant
\begin{equation}\label{3}
p_{mn}(\t,\s)=i\bar{U}\g_{mn}U=2i[u^{\a}(\s_{mn})_{\a}^{\b}u^{\b} + \bar{u}_{\dot\a}(\tilde{\s}_{mn})^{\dot\a}_{\dot\b}\bar{u}^{\dot\b}], 
\end{equation}
where $ U_{a}$  is a Majorana bispinor 
\begin{eqnarray}\label{4}
\nonumber
U_{a}=\left(\begin{array}[c]{c} u_{\a}\\ \bar{u}_{\dot\a} \end{array}\right) ,
\,\, \g_{mn}={1 \over 2}[\g_{m}, \g_{n}],\\ 
\s_{mn}={1 \over 4}(\s_{m}\tilde{\s_{n}} - \s_{n}\tilde{\s_{m}}).
\end{eqnarray}

For tensile string the bivector $ p_{mn}(\t,\s)$ may be presented as a sum of two null bivectors $p^{(+)}_{mn}$ and  $p^{(-)}_{mn}$   \cite{gz}
\begin{equation}\label{5}
p_{mn}=p^{(+)}_{mn} + p^{(-)}_{mn} =i\,[\bar{U}\g_{mn}U + \bar{V}\g_{mn}V],
\end{equation}
where $V_{a}=\left(\begin{array}[c]{c} v_{\a}\\ \bar{v}^{\dot\a} \end{array}\right)$ is  the second  component of  Newman-Penrose dyads $(u_{\a}(\t,\s), v_{\a}(\t,\s))$
\begin{equation}\label{6}
u^{\a}v_{\a}=1 ,\quad u^{\a}u_{\a}=v^{\a}v_{\a}=0
\end{equation}
and the  $\L(\t,\s)-$term is 
\begin{equation}\label{7}
 \L(\t,\s)=\l(u^{\a}v_{\a}-1) - \bar{\l}(\bar{u}^{\dot\a}\bar{v}_{\dot\a}-1).
\end{equation}

The action \f{1} may be rewritten in an equivalent spinor form
\begin{equation}\label{8}
S=i\k\int [\, p^{ab}\,d{x_{ae}}{\wedge}\,d{x_{db}}C^{ed}+ \L\, ],
\end{equation}
where $C^{ed}=(\g^0)^{ed}$  is the charge conjugation matrix in the Majorana
representation and  $p^{ab}$ is a symmetric local spin-tensor. In the general case 
 $ p^{ab}$ may be presented as a bilinear combination of the Majorana bispinors $ U_{a}$  and  $V_{a}$.
\begin{equation}\label{9}
p^{ab}=\a\, U^{a}U^{b} + \b\, V^{a}V^{b} + \r\, (U^{a}V^{b} + U^{b}V^{a})
\end{equation}
with arbitrary  coefficients  $\a,\b$ and $\r$.
 
The representation \f{8} includes an interesting object - the differential 2-form of the worldsheet area element $\xi_{a b}$ in the spinor representation
\begin{equation}\label{10}
\xi_{ab}=\xi_{ba}= C^{ed}\,d{x_{ae}}{\wedge}\,d{x_{db}},
\end{equation}
where
\begin{equation}\label{11}
dx_{ab}= (\g_{m}\,C^{-1})_{ab}\,d{x^{m}}.
\end{equation}
Unlike of the vector representation for the area element $ d{x_m}{\wedge}\,d{x_n}$ the the spinor representation  $\xi_{ab}$  is a symmetric spin-tensor 2-form under permutations of the spinor indices $a$ and  $b$.

To include the real antisymmetric central charge coordinates $z_{mn}$ we note that they may be presented by a symmetric real spin-tensor  $z_{ab}$
\begin{equation}\label{12}
z_{ab}=iz_{mn}(\g^{mn}\,C^{-1})_{ab},
\end{equation}
Then following \cite{ruds} we replace the world vector $x_{ab}$ by  a more general spin-tensor $Y_{ab}$
\begin{equation}\label{13}
x_{ab}\longrightarrow Y_{ab}={x^{m}}\,(\g_{m}\,C^{-1})_{ab}+iz_{mn}\,(\g^{mn}\,C^{-1})_{ab}.
\end{equation}
It was remarked in \cite{zli} that the differential $dY_{ab}$ \f{13} may be used as a bilding block for the consruction of a generalized differential area element $\Xi_{ab}$
\begin{equation}\label{14}
\xi_{ab}\longrightarrow \Xi_{ab}=dY_{al}\,{\wedge}{dY^{l}\,_{b}}.
\end{equation}
As a result of the extension  \f{14} the string action \f{8} is also generalized to the form,
\begin{equation}\label{15}
S=i\k\int (p^{ab}\,\Xi_{ab}+ \L\,)\\
=i\k\int (p^{ab}\,dY_{ae}\,{\wedge}dY_{db}\,C^{ed}+ \L\,).
\end{equation}
proposed in \cite{zli},  to include the TCC coordinates $z_{mn}$.
 The modification of the area element $d{x_m}{\wedge}\,d{x_n}$ induced by the contribution of $z_{lm}$ becomes more apparent after the substitution of $ Y_{ab}$  \f{13} into \f{14} which gives the following representation for the generalized area element $\Xi_{ab}$ 
\begin{equation}\label{16}
\Xi_{a}\,^{b}=(\,d{x_m}{\wedge}\,d{x_n}-8dz_{ml}{\wedge}\,dz_{n}\,^{l})(\g^{mn})_{a}\,^{b}-4i\,d{x^l}{\wedge}\,dz_{lm}(\g^{m})_{a}\,^{b}.
\end{equation}
 To derive  the representation \f{16} the standard relations for the $\g-$ matrices  
\begin{equation}\label{17}
[\g^{n_{1}n_{2}}, \g^m ]=2\,(\eta^{n_{1}m}\g^{n_2}-\eta^{n_{2}m}\g^{n_1})],
\end{equation} 
\begin{equation}\label{18}
[\g^{m_{1}m_{2}}, \g^{n_{1}n_{2}}]=2\,(\eta^{m_{1}n_{1}}\g^{m_{2}n_{2}}-\eta^{m_{2}n_{1}}\g^{m_{1}n_{2}} +  \eta^{m_{1}n_{2}}\g^{n_{1}m_{2}}-  \eta^{m_{2}n_{2}}\g^{n_{1}m_{1}}\,)
\end{equation}
have been used.

As a result the generalized action $S$ \f{15} takes the following form
\begin{equation}\label{19}
S=i\k\int \{\,[\,(\,dx_m{\wedge}\,dx_n - 8dz_{ml}{\wedge}\,dz_{n}\,^{l})\g^{mn}
-4i\,dx^l{\wedge}\,dz_{lm}\g^{m}]_{a}\,^{b}\,p^{a}\,_{b} +  \L\,\}.
\end{equation}
In \f{19}  $x^{m}$ and $z^{mn}$ appear on equal footing, and  we would  ultimately want to st0udy  solutions to this model. In a first investigation, however, we consider the minimally extended action obtained by omitting the term quadratic in $z$
\begin{equation}\label{20}
S=i\k\int \,[(\,dx_m{\wedge}\,dx_n\g^{mn}
-4i\,dx^l{\wedge}\,dz_{lm}\g^{m})_{a}\,^{b}\,p^{a}\,_{b} +  \L\,] .
 \end{equation} 

The model \f{20} was treated in \cite{zli} by considering a tensionless string minimally extended by the tensor central charge coordinates. This case corresponds to $\L=0$ and a spin-tensor $p^{ab}$ of the form  \cite{gz}
\begin{eqnarray*}
 \nonumber
p^{ab}=U^{a}U^{b}.
\end{eqnarray*}
The  minimally extended  null string action then takes the form
\begin{equation}\label{21}
S=i\int \,U^{a}\,(\,dx_m{\wedge}\,dx_{n}\g^{mn}
-4i\,dx^l{\wedge}\,dz_{lm}\g^{m})_{a}\,^{b}\,\,U_{b}\; , 
\end{equation}
where the constant $\k$ is in a redefinition of $x_{m}$ and $z_{lm}$ making all variables in \f{21} dimensionless.

In \cite{zli} a  Lagrangian treatment of the  model  \f{21} was given.
In view of the linear character of $S$ \f{20} in the $\t$ world-sheet 
derivatives  ${\dot x}_{m}$ and  ${\dot z}_{lm}$ this model is characterized by a non-trivial set of constraints and it quantization requires an investigation of these constraints and construction of the corresponding Hamiltonian mechanics.

  We shall investigate this problem below.

\section { Tensionless string with central charge coordinates. The primary constraints}

To study the constraints and Hamiltonian dynamics of  $S$  \f{21} we start 
from its  Weyl representation \cite{zli}
\begin{eqnarray}\label{22}
S=i\int[ (u^{\a}dx_{\a\dot\l}{\wedge}d\tilde x^{\dot\l\b}u_{\b}+ \bar{u}_{\dot\a}d\tilde {x}^{\dot{\a}\l}{\wedge}dx_{\l\dot\b}\bar{u}^{\dot\b})
\nonumber\\ 
+  2 (u_{\a}dz^{\a\b}{\wedge}dx_{\b\dot\l}\bar{u}^{\dot\l} - \bar{u}_{\dot\a}d\tilde z^{\dot\a\dot\b}\wedge u^{\l}dx_{\l\dot\b}) ], 
\end{eqnarray}
To make the transition to the Weyl basis we used the following relations
\begin{eqnarray}\label{23}
{\S_\a}^\b\equiv d{x_m}{\wedge}\,d{x_n}(\s^{mn})_{\a}\,^{\b}={1\over 2}\,
dx_{\a\dot\l}\wedge\,d\tilde x^{\dot\l\b},\nonumber \\
\tilde{\S}^{\dot\b}\,_{\dot\a}\equiv -({\S_\a}^\b)^\ast =
d{x_m}\wedge\,dx_{n}(\tilde{\s}^{mn})^{\dot\b}\,_{\dot\a}={1\over 2}\,d\tilde
{x}^{\dot\b\l}\wedge dx_{\l\dot\a},
\end{eqnarray}
where
\begin{eqnarray}\label{24}
\nonumber 
dx_{\a\dot\b}=(\s_{m})_{\a \dot\b}d{x^m}, \,\,\,
d{x^m}=-{1\over 2}(\tilde{\s}^{m})^{\dot\a\b}dx_{\b\dot\a},\\
((\s^{mn})_{\a}\,^{\b}))^{\ast}=-(\tilde{\s}^{mn})^{\dot\b}\,_{\dot\a}, \,\,\,
d\tilde{x}^{\dot\a\b}=\e^{\dot\a\dot\l}\e^{\b\g}dx_{\g\dot\l},
\end{eqnarray}
and the antihermitian 2-form $\Om_{\a\dot\b}$ 
\begin{eqnarray}\label{25}
\Om_{\a\dot\b} \equiv -8i(dz_{ml}\wedge\,dx^{l})(\s^m)_{\a\dot\b}=2[(dz_{\a}\,^{\l}\wedge\,dx_{\l\dot\b}+dx_{\a\dot\l} d\bar z^{\dot\l}\,_{\dot\b}] , \nonumber \\
(\Om_{\a\dot\b})^\ast=-\Om_{\b\dot\a},
\end{eqnarray}
where
\begin{equation}\label{26}
z_{m_{1}m_{2}}= {i\over 4}[\,z_{\a}\,^{\b}(\s_{m_{1}m_2})_{\b}\,^{\a} +  \bar{z}_{\dot\a}\,^{\dot\b}(\tilde{\s}_{m_{1}m_2})^{\dot\a}\,_{\dot\b}\,]
\end{equation}
Note that  $S$ \f{22} may be equivalently presented in the compact form  
\begin{equation}\label{27}
S=2i\int( u^{\a}{\S_\a}^{\b}u_{\b}+ \bar{u}_{\dot\b}\tilde{\S}^{\dot\b}\,_{\dot\a} \bar{u}^{\dot\a}+{1\over 2} u^{\a}\Om_{\a\dot\b}\bar{u}^{\dot\b}).
\end{equation}

For the Hamiltonian description we need the canonical  momentum densities
\begin{equation}\label{28}
P^{\cal M}\equiv(P^{\a\dot\a},\pi^{\a\b},{\bar \pi}^{\dot\a\dot\b},
P_{u}^{\a}, \bar{P}_{u}^{\dot\a}, P_{v}^{\a}, \bar{P}_{v}^{\dot\a})=
{\partial {\cal L}\over \partial \dot{q}_{\cal M}}
\end{equation}
which are conjugate to the target space coordinates
\begin{equation}\label{29}
q_{\cal M}=( x_{\a\dot\a},z_{\a\b},{\bar z}_{\dot\a\dot\b},
u_\a, \bar u_{\dot\a}, v_\a, \bar v_{\dot\a} )
\end{equation}
in the Poisson bracket
\begin{equation}\label{30}
\{P^{\cal M}(\s), q_{\cal N}(\s^\prime)\}_{P.B.}=\d^{\cal M}_{\cal N}\d(\s -\s^\prime) 
\end{equation}
As far as  the action  \f{22} is  linear  in the  proper time derivative 
$\dot{x}_{\a\dot\a}, \dot{z}_{\a\b}, \dot{\bar{z}}_{\dot\a\dot\b}$ the definition \f{28} leads to the  primary constraints $\Phi$  
\begin{equation}\label{31}
\Phi^{\a\dot\a}\equiv(P^{\a\dot\a} - \triangle_{\Phi}^{\a\dot\a})=0,
\end{equation}
where\begin{eqnarray}\label{32}
\nonumber 
\triangle_{\Phi}^{\a\dot\a}\equiv2i( u^{\a}\bar{r}^{\dot\a}-r^{\a}\bar{u}^{\dot\a}),\\
r^\a \equiv(\bar{u}\tilde{x}^\prime)^\a + (uz^\prime)^\a,
\end{eqnarray}
and $\Psi$
\begin{equation}\label{33}
\Psi^{\a\b}\equiv\pi^{\a\b} + \triangle_{\Psi}^{\a\b}=0,
\end{equation}
where
\begin{equation}\label{34}
\triangle_{\Psi}^{\a\b}\equiv2i[u^{\a}(\bar{u}\tilde{x}^\prime)^\b  +
u^{\b}(\bar{u}\tilde{x}^\prime)^\a] .
\end{equation}
 These constraints have to be added by the  primary constraints for the dyads $u^\a$ and $ v^\a$  
\begin{equation}\label{35}
P_{u}^{\a}=0, \;\; P_{v}^{\a}=0,
\end{equation}
together with their complex conjugate (c.c.).
 
The standard definition of the canonical Hamiltonian density 
\begin{equation}\label{36}
{\cal H}_{0}= {\cal P}^{{\cal M}^{\prime}}\dot{q}_{{\cal M}^{\prime}} + {1\over 2}\pi^{\a\b}\dot{z}_{\a\b} + {1\over 2}{\bar \pi}^{\dot\a\dot\b}\dot{\bar z}_{\dot\a\dot\b}- \cal L,
\end{equation}
where
\begin{equation}\label{37}
q_{\cal M^{\prime}}\equiv(x_{\a\dot\a}, u_\a, \bar u_{\dot\a}, v_\a, \bar v_{\dot\a} ),
\end{equation}
 is consistent with the Poisson brackets  definition \f{30} and generates the Hamiltonian equations  of motion   
\begin{equation}\label{38}
{df\over d\t}\equiv\dot f(\t,\s) =\int d{\s^{\prime}}\{{\cal H}_{0}(\s^{\prime}),\,  f(\t,\s)\}_{P.B.}.
\end{equation}

 In the considered case we have from the definitions \f{28} and  \f{36} that
\begin{equation}\label{39}
{\cal H}_{0}=0
\end{equation}
and the evolution of the  string is described  by the  generalized Hamiltonian
\begin{eqnarray}\label{40}
\nonumber 
{\cal H}={\cal H}_{1} + {\cal H}_{2}, \\
\nonumber
{\cal H}_{1}\equiv a_{\a\dot\a}\Phi^{\a\dot\a} +  b_{\a\b}\Psi^{\a\b}+
 {\bar b}_{\dot\a\dot\b}{\bar \Psi}^{\dot\a \dot\b},
\nonumber\\
{\cal H}_{2}=[\mu_{\a}P_{u}^\a +\r_{\a}P_{v}^\a + \l(u^{\a}v_{\a} - 1)] + c.c.
\end{eqnarray} 

Before we embark on the  analysis of the consistency  of the constraints,  we remark on  some interesting consequences  of \f{31} and \f{33}, e.g., 
\begin{equation}\label{41}
P^{\dot\a\a}P_{\a\dot\a}+ 8(ux^{\prime}\bar u + uz^{\prime}u)
(ux^{\prime}\bar u+{\bar u}{\bar z}^{\prime}\bar u)=0
\end{equation}
and 
\begin{equation}\label{42}
P^{\dot\a\a}x^{\prime}_{\a\dot\a}=2i(ux^{\prime}z^{\prime}\bar u  - ux^{\prime} z^{\prime}\bar u),
\end{equation}

\begin{equation}\label{43}
\pi^{\a\b}\pi_{\a\b}={\bar\pi}^{\a\b}{\bar\pi}_{\dot\a\dot\b}=4(ux^{\prime}{\bar u})^2 ,
\end{equation}

\begin{equation}\label{44}
\pi^{\a\b}z^{\prime}_{\a\b}=4i(uz^{\prime}x^{\prime} \bar u),\,\,
 {\bar \pi}^{\dot\a\dot\b}{\bar z}^{\prime}_{\dot\a\dot\b}=-4i(ux^{\prime}z^{\prime}\bar u),
\end{equation}
It follows  from \f{41}, \f{42} and \f{43} that  the string  constraints reduce  to the tensionless string  constraints  \cite{ulf2,lizzi}
\begin{equation}\label{45}
P^{\dot\a\a}P_{\a\dot\a}{\mid}_{z=\pi=0}=0, \;\;\;  P^{\dot\a\a}x^{\prime}_{\a\dot\a}|_{z=\pi=0}=0,
 \end{equation}
if the TCC coordinates $ z_{\a\b}$ and their momenta $ \pi^{\a\b}$ are equal zero. It shows a nontrivaial contribution of the  TCC coordinates in the constraint algebra and the null string dynamics. 
Moreover, the constraint  \f{42} may be  presented in a slightly different form
\begin{eqnarray}\label{46}
\nonumber 
P^{\dot\a\a}x^{\prime}_{\a\dot\a} +{1\over 2}( \pi^{\a\b}z^{\prime}_{\a\b}
+{\bar \pi}^{\dot\a\dot\b}{\bar z}^{\prime}_{\dot\a\dot\b})\\
\equiv\Phi^{\dot\a\a}x^{\prime}_{\a\dot\a}+{1\over 2}( \Psi^{\a\b}z^{\prime}_{\a\b}
+{\bar \Psi}^{\dot\a\dot\b}{\bar z}^{\prime}_{\dot\a\dot\b})=0
\end{eqnarray}
using  \f{44}.
The constraint  \f{46} together with the constraints \f{35} are equivalent to  to the first class constraints $T_{\pm}$
\begin{eqnarray}\label{47}
\nonumber 
T_{\pm}{\equiv}P^{\dot\a\a}x^{\prime}_{\a\dot\a} +{1\over 2}( \pi^{\a\b}z^{\prime}_{\a\b} + {\bar \pi}^{\dot\a\dot\b}{\bar z}^{\prime}_{\dot\a\dot\b})\\
\pm [(u_{\a}^{\prime}P_{u}^\a +{\bar u }^{\prime}_{\dot\a}{\bar P}_{u}^{\dot\a}) + (v_{\a}^{\prime}P_{v}^\a + {\bar v }^{\prime}_{\dot\a}{\bar P}_{v}^{\dot\a}  )]= 0,
\end{eqnarray}
where  $T_{\pm}$ describe the world sheet  diffeomorphisms corresponding to 
 $\s$-shifts (modulo the constraints \f{35} for the sign  $(-)$ in \f{47}).

In the next Sections we will construct the remaining first and second class constraints using the Dirac's selfconsistency procedure.

\section  {The first class constraints}

The conservation conditions of the  constraints  \f{6}, \f{31}, \f{33}, 
\f{35} and their c.c. 
\begin{equation}\label{48}
{d{\phi}\over d\t}\equiv\dot\phi(\t,\s) =\int d{\s^{\prime}}\{{\cal H}(\s^{\prime}),\, \phi(\t,\s)\}_{P.B.}\approx 0,
\end{equation}
where the symbol ${\approx} $ means the Dirac's weak equality, must either restrict the Lagrange multipliers in {\cal H} \f{40} or/and to produce new secondary constraints.  The conservation of the  constraint \f{6} leads to the condition 
\begin{equation}\label{49}
\r_{\a}u^{\a} -\mu_{\a}v^{\a}=0,
\end{equation}which has  the general solution 
\begin{equation}\label{50}
\mu_{\a}=hu_{\a}+ gv_{\a},\;\;
\r_\a =-hv_{\a} + \nu{v_\a}.
\end{equation}
Analogously, the  conservation condition for the  constraint $P_{v}^{\a}=0$  
 \f{35} gives 
\begin{equation}\label{51}
\l=0 .
\end{equation}
As a result ${\cal H}_2$ in \f{40} takes the form 
\begin{equation}\label{52}
{\cal H}_{2}=[h(u_{\a}P_{u}^{\a}- v_{\a}P_{v}^{\a})+gv_{\a}P_{u}^{\a} + {\nu} u_{\a}P_{v}^{\a}] + c.c.
\end{equation}
It is easy to check that 
\begin{equation}\label{53}
W\equiv u_{\a}P_{v}^{\a}=0
\end{equation}
and its c.c.  are the  first class constraints generating the gauge symmetry of the action by  the complex shifts
\begin{equation}\label{54}
{\d}v_{\a}={\d\nu}u_{\a},\,\; \; {\d}u_{\a}=0.
\end{equation}

To check the  conservation of the  constraint $P_{u}^{\a}=0$ we note that 
\begin{eqnarray}\label{55}
\nonumber 
\{P_{u}^{\g}(\s), \Phi^{\a\dot\a}(\s^\prime)\}_{P.B.}=
-2i[ u^\a{\tilde x}^{\prime \dot\a\g} - {\bar u}^{\dot\a}z^{\prime\a\g} + \e^{\a\g}({\tilde x}^\prime u + {\bar u}{\bar z}^\prime )^{\dot\a} ]\d(\s-\s^\prime),\\
\nonumber 
\{P_{u}^{\g}(\s), \Psi^{\a\b}(\s^\prime)\}_{P.B.}=2i[\e^{\a\g}({\bar u}{\tilde x}^{\prime})^{\b} +\e^{\b\g}({\bar u}{\tilde x}^{\prime})^{\a} ]\d(\s-\s^\prime),\\
\{P_{u}^{\g}(\s), {\bar\Psi}^{\dot\a\dot\b}(\s^\prime)\}_{P.B.}=-2i[{\bar u}^{\dot\a}\tilde{x}^{\prime\dot\b\g } +  {\bar u}^{\dot\b}\tilde{x}^{\prime\dot\a\g}]\d(\s-\s^\prime).
\end{eqnarray}
Then we find the following equation
\begin{equation}\label{56}
{\dot P}_{u}^{\g}=2i\,[u(a{\tilde x}^\prime - x^{\prime}\tilde a)
 -(z^{\prime}a-2bx^{\prime})\bar u -{\bar u}({\bar z}^{\prime}\tilde{a}
-2{\bar b}{\tilde x}^{\prime})]^{\g}\approx 0
\end{equation}
which does not involve the  Lagrange  multipliers $h, g $ and $\nu$. It follows from  \f{47} that  Eqs. \f{56} have as a particular solution 
\begin{equation}\label{57}
a_{\a\dot\a}=a_{0} x^{\prime}_{\a\dot\a}, \; b_{\a\b}=b_{0} z^{\prime}_{\a\b}, \;a_{0}=\bar a_{0}, \; b_{0}=\bar b_{0}.
\end{equation}
In fact, substituting \f{57} into \f{56} transforms the latter into the equation 
\begin{equation}\label{58}
{\dot P}_{u}^{\g}=-2i\,(a_{0} -2b_{0})(z^{\prime}x^{\prime}\bar u +{\bar u}{\bar z}^{\prime}{\tilde x}^{\prime})^{\g}\approx 0 ,
\end{equation}
which  has the expected  solution
\begin{equation}\label{59}
 b_{0}={1\over 2}\,a_{0}.
\end{equation}

To find the second solution of \f{56} and  the solutions for the Lagrange multipliers $h, g $ and $\nu$ from  ${\cal H}_2$ we consider the consistency conditions for $\Phi^{\a\dot\a}$ and $\Psi^{\a\b}$.

 To this end we  note that 
\begin{eqnarray}\label{60}
\nonumber 
\{\Phi^{\a\dot\a}(\s), \Phi^{\b\dot\b}(\s^\prime)\}_{P.B.}=2i[( \e^{\a\b}{\bar u}^{\dot\a}{\bar u}^{\dot\b} -  u^{\a}u^{\b} \e^{\dot\a\dot\b})|_{\s^\prime}\\
\nonumber   -( \s^\prime \rightarrow  \s)]\partial_{\s}\d(\s-\s^\prime),\\
\{\Psi^{\a\b}(\s), \Psi^{\g\d}(\s^\prime)\}_{P.B.}=
\{\Psi^{\a\b}(\s), {\bar \Psi}^{\dot\g\dot\d}(\s^\prime)\}_{P.B.}=0
\end{eqnarray}
and 
\begin{equation}\label{61}
\{\Phi^{\a\dot\a}(\s), \Psi^{\b\g}(\s^\prime)\}_{P.B.}=-2i[(\e^{\a\b}u^{\g} + \e^{\a\g}u^{\b}){\bar u}^{\dot \a}|_\s -( \s^\prime \rightarrow  \s)]\partial_{\s}\d(\s-\s^\prime).
\end{equation}
Using  \f{60} and \f{61} we find 
\begin{eqnarray}\label{62}
\nonumber 
{1\over 2i}{\dot \Psi}^{\b\g}=[(u^{\prime\b}({\bar u}{\tilde a})^\g +
 u^{\b}({\bar u}^{\prime}{\tilde a})^{\g}) + (\b \rightarrow \g) ]\\
+ [({\zeta}^{\b}({\bar u}{\tilde x}^{\prime})^\g +
u^{\b}({\bar\zeta}{\tilde x}^{\prime})^{\g}) + (\b \rightarrow \g)],
\end{eqnarray}
where
\begin{equation}\label{63}
\zeta^{\a}=hu^\a + gv^\a .
\end{equation}
After multiplication \f{62} by $u_{\b}u_\g $ we find 
\begin{equation}\label{64}
g=(u^{\prime\b}u_{\b}){({\bar u}{\tilde a}u)\over ({\bar u}{\tilde x}^{\prime}u)}.
\end{equation}
Moreover, the  $u^{\prime}_\a$ expansion in the base dyades  $u_{\a}$ and 
$v_{\a}$
\begin{equation}\label{65}
u^{\prime}_\a =\varphi u_{\a} + \chi v_{\a}
\end{equation}
transforms equations \f{62} and \f{64} into the equations 
\begin{eqnarray}\label{66}
\nonumber
\{
 u^\b[
  (\varphi + \bar\varphi)({\bar u}{\tilde a})^\g +
  (h+\bar h)({\bar u}{\tilde x}^{\prime})^\g +
  \bar\chi({\bar v}{\tilde a})^\g +
  \bar g ({\bar v}{\tilde x}^{\prime})^\g
 ] +v^\b[\chi({\bar u}{\tilde a})^\g + g({\bar u}{\tilde x}^{\prime})^\g
 ]
\}  \\
+ \, \{ \b \rightarrow \g
\}=0 
\end{eqnarray}
and 
\begin{equation}\label{67}
g=-\chi{({\bar u}{\tilde a}u)\over ({\bar u}{\tilde x}^\prime u)} .
\end{equation}
In the first place  we note that the particular solution \f{57} for
${\tilde a}^{\dot\a\a}$
\begin{equation}\label{68}
{\tilde a}^{\dot\a\a}=a_0{\tilde x}^{\prime\dot\a\a}, \;\;\; a_0= {\bar a}_0
\end{equation}
have to be a solution of Eqs. \f{66} and \f{67}. After a  substitution of 
\f{68} into  these equations they are reduced to the relations 
\begin{equation}\label{69}
(a_{0}\varphi + h) +({\bar a}_{0}\bar\varphi +\bar h),\; \;g=-a_0{\chi}
\end{equation}
or equivalently to
\begin{eqnarray}\label{70}
\nonumber
h=-a_{0}\varphi  + i\th_0, \,  \;g=-a_{0}\chi ,
\\
\varphi=(u^{\prime\a}v_{\a}), \; \; \chi=-(u^{\prime\a}u_{\a}), 
\end{eqnarray}
where $\th_0$ is an arbitrary real function. The expression  for ${\cal H}_2$
\f{52} corresponding  to the solution \f{70} takes the form
\begin{equation}\label{71}
{\cal H}_{2}=[-a_{0}(u^{\prime}_{\a}P_{u}^{\a} + v^{\prime}_{\a}P_{v}^{\a})
 + {\tilde\nu} u_{\a}P_{v}^{\a} +i\th_{0}(u_{\a}P_{u}^{\a}-v_{\a}P_{v}^{\a})] + c.c.,
\end{equation}
where  $\tilde\nu \equiv\nu +(v^{\prime\a}v_{\a})$
 and the relation 
\begin{equation}\label{72}
u_{\a}v_{\b}- v_{\a}u_{\b}=\e_{\a\b}
\end{equation}
was used. The expression for  ${\cal H}_2$ given in \f{71} and  that for   ${\cal H}_1$ in \f{40} shows that the Lagrangian multiplier $a_0$ corresponds to the first class constraints  $T_{-}$ (which equals  $T_{+}$  modulo the
constraints  $P_{u}^{\a}\approx 0$ and $P_{v}^{\a}\approx 0$).
The last $\th_0$-term in \f{71} shows the invariance of $\Psi^{\a\b}$ under the transformations 
\begin{equation}\label{73}
u^{\prime}_{\a}=e^{i\Th(\t,\s)}{u}_{\a},\; \; v^{\prime}_{\a}=e^{-i\Th(\t,\s)}{v}_{\a}.
\end{equation}
However, the transformations \f{73} is not a symmetry of the action \f{22}  and consequently we have to choose the integration  ``constant''  $\th_0$ in the  solution  \f{70} vanish
\begin{equation}\label{74}
\th_{0}(\t,\s)=0.
\end{equation}

We conclude that the solution  \f{68} generates the first class constraint 
\begin{eqnarray}\label{75}
\nonumber 
T{\equiv}P^{\dot\a\a}x^{\prime}_{\a\dot\a} +{1\over 2}( \pi^{\a\b}z^{\prime}_{\a\b} + {\bar \pi}^{\dot\a\dot\b}{\bar z}^{\prime}_{\dot\a\dot\b})\\
+ [(u_{\a}^{\prime}P_{u}^\a +{\bar u }^{\prime}_{\dot\a}{\bar P}_{u}^{\dot\a}) + (v_{\a}^{\prime}P_{v}^\a + {\bar v }^{\prime}_{\dot\a}{\bar P}_{v}^{\dot\a}  )]\approx 0
\end{eqnarray}
which is one of the generators of the generalized Virasoro algebra 
corresponding to the $\s-$reparametrization of the string world sheet.

The second generator of this algebra corresponds to the second  solution of 
Eqs. \f{66} and \f{67}
\begin{equation}\label{76}
{\tilde a}^{\dot\a\a}=a_{1}u^{\a}{\bar u}^{\dot\a}, \; \; a_1= {\bar a}_1
\end{equation}
which restricts the corresponding Lagrange multipliers $ g, h $ and $ \chi$ in  ${\cal H}_2$  \f{52} to be 
\begin{equation}\label{77}
g=0, \;\; h=i\th_1, \;\; \chi=0,
\end{equation}
where $\th_1 (\t,\s)$ is an arbitrary real function. Due to the general  
$u^{\prime}_\a$ expansion  \f{65}  we conclude that the solution  \f{77} implies a  secondary constraint
\begin{equation}\label{78}
u^{\prime\a}u_\a =0.
\end{equation}
It is easy to check that this constraint is preserved by the Hamiltonian \f{40} because 
\begin{equation}\label{79}
g=0
\end{equation}
for the both solutions \f{70} and \f{77}.  The constraint  \f{78} is a second class constraint. 

The solutions  \f{68}  and \f{76} have also to be the solutions of the equation 
\begin{equation}\label{80}
\dot\Phi^{\dot\a\a}(\t,\s) =\int d{\s^{\prime}}\{{\cal H}(\s^{\prime}),\,
 \Phi^{\a\dot\a}(\t,\s)\}_{P.B.}\approx 0,
\end{equation}
Using the P.B. \f{60} and \f{61} we find 
\begin{eqnarray}\label{81}
\nonumber 
\{{\cal H}_1 + {\cal H}_2 , \Phi^{{\dot\a}\a}\}_{P.B.}=-2i[u^{\prime\a}
 ({\tilde a}u)^{\dot\a}+ 
u^{\a}({\tilde a}u^{\prime})^{\dot{\a}}  \\
+ 2(u^{\prime\a}({\bar b}{\bar u})^{\dot\a} + u^{\a}({\bar b}{\bar u}^\prime)^{\dot\a}) + {\zeta}^\a ({\tilde x}^\prime u )^{\dot\a} +  
{u}^{\a}({\tilde x}^{\prime}{\zeta})^{\dot\a} +   
{\zeta}^{\a}({\bar z}^{\prime}{\bar u})^{\dot\a}+ {u}^{\a}({\bar z}^{\prime}{\bar\zeta})^{\dot\a}] - c.c.\approx 0.                               
\end{eqnarray}
The substitution of the expansion \f{65} 
\begin{equation}\label{82}
{\zeta}_\a= hu_\a , \;\; u_{\a}^{\prime}=\varphi{u_\a} 
\end{equation}
into Eqs. \f{81} and using the solution \f{79} leads  to                       \begin{equation}\label{83}
\{2[\varphi({\tilde a}u)^{\dot\a}+h({\tilde x}^{\prime}u)^{\dot\a}]
+ [2(\varphi +\bar\varphi)({\bar b}{\bar u})^{\dot\a}+ (h+\bar h)({\bar z}^{\prime}{\bar u})^{\dot\a}]\}-c.c\approx 0.
\end{equation}
The substitution of \f{57} and \f{59} into \f{83} yields the equation
\begin{equation}\label{84}
\{2(a_{0}\varphi +h)({\tilde x}^{\prime}u)^{\dot\a}+
 [a_{0}(\varphi +\bar\varphi)+ (h+\bar h)]({\bar z}^{\prime}{\bar u})^{\dot\a}]\}-c.c\approx 0
\end{equation}
which is satisfied due to the relations \f{70} and \f{74}.

The substitution of the second solution \f{76} and \f{77} into Eq. \f{83} gives  the equation 
\begin{equation}\label{85}
[h({\tilde x}^{\prime}u)^{\dot\a} + (\varphi +\bar\varphi)({\bar b}{\bar u})^{\dot\a}]- c.c.\approx 0
\end{equation}
which has the following solution for the Lagrange multipliers $h$ and $b_{\a\b}$ 
\begin{equation}\label{86}
h=0, \;\; b_{\a\b}=b_{1}u_{\a}u_\b .
\end{equation}

The next step is the substitution of the solutions \f{76} and \f{86} into 
Eq. \f{56} which  transform it into the final equation 
\begin{equation}\label{87}
a_{1}[(ux^{\prime}{\bar u}) +  ({\bar u}{\bar z}^{\prime}{\bar u})]=2b_{1}(ux^{\prime}{\bar u}).
\end{equation}
As long  as $(ux^{\prime}{\bar u})\neq 0$,  Eq. \f{76} allows us to express the  Lagrange multiplier $b_{1}$ as a function of $a_{1}$ 
\begin{equation}\label{88}
b_{1}= {a_{1}\over 2}[1+{({\bar u}{\bar z}^{\prime}{\bar u})\over (ux^{\prime}{\bar u})}],\;\;
{\bar b_{1}}= {a_{1}\over 2}[1+{(u{z}^{\prime}u)\over (ux^{\prime}\bar u)}].
\end{equation}
Using Eqs.\f{76}, \f{77}, \f{79}, \f{64} and the expression for $ {\cal H}_2$  we find the Hamiltonian density \f{40} corresponding  to the second solution of the selfconsistency conditions
\begin{eqnarray}\label{89}
\nonumber  
{\cal H}= a_{1}\{u_{\a}{\bar u}_{\dot\a}\Phi^{\dot\a\a}
+{1\over 2}[1+{({\bar u}{\bar z}^{\prime}{\bar u})\over (ux^{\prime}{\bar u})}]u_{\a}u_{\b }\Psi^{\a\b}   \nonumber\\
+{1\over 2}[1+{(u{z}^{\prime}u)\over (ux^{\prime}{\bar u})}]{\bar u}_{\dot\a}{\bar u}_{\dot\b}{\bar \Psi}^{\dot\a\dot\b}\} +[i\th_{1}(u_{\a}P_{u}^{\a}-v_{\a}P_{v}^{\a}) + c.c.].
\end{eqnarray}
In just the same way as above we resume that the integration ``constant'' $\th_{1}$ should  vanish
\begin{equation}\label{90}
\th_{1}(\t,\s)=0,
\end{equation}
because it corresponds to the transformations \f{73} which are not symmetries of the  action \f{22}. 
Then ${\cal H}$ in \f{89} yields the second Virasoro generator 
\begin{equation}\label{91}
Q\equiv P^{\dot\a\a}{u_{\a}{\bar u}_{\dot\a}+ {1\over 2}[1+{({\bar u}{\bar z}^{\prime}{\bar u})\over (ux^{\prime}{\bar u})}]{\pi}^{\a\b}u_{\a}u_{\b } +
{1\over 2}[1+{(u{z}^{\prime}u)\over (ux^{\prime}{\bar u})}]}{\bar \pi}^{\dot\a\dot\b}{\bar u}_{\dot\a}{\bar u}_{\dot\b} \approx 0 .
\end{equation}
This result completes the Dirac procedure for the construction of the first class constraints $W$ (eqn. \f{53}),  $T$ (eqn. \f{75}) ,  $Q$ (eqn. \f{91}). 

\section {Hamiltonian and algebra of the first class constraints}

In the previous Section we found four real first class constraints:  two real constraints $T$ and $Q$ and one complex constraint $W$
\begin{eqnarray}\label{92}
\nonumber  
T{\equiv}P^{\dot\a\a}x^{\prime}_{\a\dot\a} +{1\over 2}( \pi^{\a\b}z^{\prime}_{\a\b} + {\bar \pi}^{\dot\a\dot\b}{\bar z}^{\prime}_{\dot\a\dot\b})\\
\nonumber  
+ (u_{\a}^{\prime}P_{u}^\a + v_{\a}^{\prime}P_{v}^\a ) +
({\bar u }^{\prime}_{\dot\a}{\bar P}_{u}^{\dot\a}
 + {\bar v }^{\prime}_{\dot\a}{\bar P}_{v}^{\dot\a} )\approx 0,\\ 
\nonumber  
Q\equiv P^{\dot\a\a}{u_{\a}{\bar u}_{\dot\a}+ {1\over 2}[1+{({\bar u}{\bar z}^{\prime}{\bar u})\over (u{\tilde x}^{\prime}{\bar u})}]{\pi}^{\a\b}u_{\a}u_{\b } +
{1\over 2}[1+{(u{z}^{\prime}u)\over (ux^{\prime}{\bar u})}]}{\bar \pi}^{\dot\a\dot\b}{\bar u}_{\dot\a}{\bar u}_{\dot\b} \approx 0 ,\\
W\equiv (u_{\a}P_{v}^{\a})\approx 0, \;\;\;\;
\bar W\equiv ({\bar u}_{\dot\a}{\bar P}_{v}^{\dot\a})\approx 0. 
\end{eqnarray}

The Hamiltonian density  ${\cal H}$ is correspondingly
 \begin{equation}\label{93}
{\cal H}=a _{0}T + a _{1}Q +  \nu W + \bar\nu{\bar W}\approx 0.
\end{equation}
If we use the following condenced  notation for the constraints $Q$, $W$ and ${\bar W}$
\begin{equation}\label{94}
{\cal K}_{I}\equiv(Q, W,\bar W ),\;\; (I=1,2.3),
\end{equation}
we find that the P.B. algebra of the  first class constraints $T$ and  ${\cal K}_{I}$ takes the compact form 
\begin{eqnarray}\label{95}
\nonumber 
\{T(\s), T(\s^\prime)\}_{P.B.}=[T(\s) +  T(\s^\prime)]\partial_{\s^\prime}(\s -\s^\prime), \\ \nonumber 
\{T(\s), {\cal K}_{I}(\s^\prime)\}_{P.B.}={\cal K}_{I}(\s)\partial_{\s^\prime}(\s -\s^\prime), \\ 
\{{\cal K}_{I}(\s), {\cal K}_{J}(\s^\prime)\}_{P.B.}=0.    
\end{eqnarray}
The algebra  coincides with the corresponding algebra for tensionless string \cite{guliss}
\begin{eqnarray}\label{96}
\nonumber 
\{T_{0}(\s), T_{0}(\s^\prime)\}_{P.B.}=[T_{0}(\s) +  T_{0}(\s^\prime)]\partial_{\s^\prime}(\s -\s^\prime), \\ \nonumber 
\{T_{0}(\s), {\cal K}_{0I}(\s^\prime)\}_{P.B.}={\cal K}_{0I}(\s)\partial_{\s^\prime}(\s -\s^\prime), \\ 
\{{\cal K}_{0I}(\s), {\cal K}_{0J}(\s^\prime)\}_{P.B.}=0,    
\end{eqnarray}
where
 \begin{equation}\label{97}
T_{0}\equiv T|_{z=\pi=0}, \;\; \;\; {\cal K}_{0I}\equiv{\cal K}_{I}|_{z=\pi=0} 
\end{equation}
So, we conclude that the minimal inclusion of the TCC coordinates $z_{\a\b}$ has no effects on the level of P.B. algebra for the first class constraint. 
To find these effects we have to make the transition to the Dirac brackets which take into account the second class constraints for the considered action
\f{27}. These constraints differ from the second class constraint of tensionless string because of the  TCC coordinates $z_{\a\b}$ presense.
 
 Note also that the structure of the P.B. algebra \f{95} resembles  the structure of the contracted rotation algebra in (anti) de Sitter space  with the splitted  ``fifth'' coordinate  and  the space radius $R$ going to infinity. 
 In view of an universal character of the $T$ generator we do not expect that the two first P.B. in the algebra \f{95} will change after the transition to the Dirac bracket. But, we may expect the  appearance of a non-zero contribution into the right hand side of the Dirac bracket originated from the last 
P.B. in \f{95}. If so, that scenario gives a hint that the introduction of  TCC coordinates  restors  a finite value of $R$ without breaking the conformal symmetry of the string action. Below we derive the second class constraints and describe their algebraic structure. We will show that the introduction of the TCC coordinates $z_{mn}$ effectively adds only one coordinate to the $x_m$ coordinates.                           
\section {The second class constraints}                                        

For the  covariant splitting of the second class constraints from the constraints  \f{31}, \f{33}, \f{35}  note that the constraints $T$ and $Q$ are the projections of $\Phi^{\dot\a\a}$ onto the 4-vectors  
$u_{\a}{\bar u}_{\dot\a}$ and $x_{\a\dot\a}^{\prime}$. So, taking into account  that  $\Phi^{\dot\a\a}$ is a 4-vector it is useful to introduce a local moving frame composed of the real 4-vectors \cite{z5,gz}                 \begin{eqnarray}\label{98}
\nonumber                                                                       n_{\a\dot\a}^{(+)}=u_{\a}{\bar u}_{\dot\a}, \;\;
 n_{\a\dot\a}^{(-)}=v_{\a}{\bar v}_{\dot\a}, \\
m_{\a\dot\a}^{(+)}=u_{\a}{\bar v}_{\dot\a}+ v_{\a}{\bar u}_{\dot\a}, 
m_{\a\dot\a}^{(-)}=i(u_{\a}{\bar v}_{\dot\a}+v_{\a}{\bar u}_{\dot\a}) 
\end{eqnarray}
 attached to the string world sheet. The 4-vector $\Phi^{\dot\a\a}$ may then be expanded in this  4-vector basis \f{98}. Taking into account the condition $(ux^{\prime}{\bar u})\neq0$,  which shows that  4-vector $x_{\a\dot\a}^{\prime}$
has a nonzero projection onto the light-like  basis  4-vector $n_{\a\dot\a}^{(-)}$, one can choose the projections
\begin{equation}\label{99}
M^{(\pm)}\equiv\Phi^{\dot\a\a}m_{\a\dot\a}^{(\pm)}=0
\end{equation}
to be the second class constraints. The primary constraints  $\Phi^{\dot\a\a}$  then split covariantly into  two real first class constraints $T,\,Q$  and two real second class constraints  $M^{\pm}$
\begin{equation}\label{100}
 \Phi^{\dot\a\a}\Rightarrow(T,\;Q)\oplus (M^{(+)}, \;M^{(-)})
\end{equation}

Taking into account  the  P.B. relations for $\Phi^{\dot\a\a}$ \f{60} together with  the ``orthonormality'' conditions 
\begin{eqnarray}\label{101}
\nonumber             
n^{(\pm)\a\dot\a}n^{(\pm)}_{\a\dot\a}=0,\;\; n^{(-)\a\dot\a}n_{\a\dot\a}^{(+)}=1,\\
\nonumber  
m^{(\pm)\a\dot\a}m_{\a\dot\a}^{(\pm)}=-2,\;\;
 m^{(-)\a\dot\a}m_{\a\dot\a}^{(+)}=0,\\
n^{(\pm)\a\dot\a}m_{\a\dot\a}^{(\pm)}=0,\;\; m^{(-)\a\dot\a}m_{\a\dot\a}^{(+)}=0,
\end{eqnarray}
 it is easy to find  that the $ M^(\pm)$-constraints have  zero P.B.
\begin{equation}\label{102}
\{M^{(\pm)}, \; M^{(\pm)}\}_{P.B.}=0, \;\;  \{M^{(+)}, \; M^{(-)}\}_{P.B.}=0
\end{equation}
It is suitable to introduce one complex constraint $M$  instead of
 two real $ M^{(\pm)}$-constraints
\begin{eqnarray}\label{103}
\nonumber  
M\equiv {1\over 2}(M^{(+)}+iM^{(-)}); \;\;
\bar M\equiv{1\over 2}((M^{(+)}- iM^{(-)})\\
\end{eqnarray}
 forming the abelian P.B. algebra
\begin{equation}\label{104}
\{M, \; M\}_{P.B.}=0, \;\;  \{M, \; \bar M\}_{P.B.}=0.
\end{equation}
As a result the primary constraint  $\Phi^{\dot\a\a}$ \f{60} are splitted into
 \begin{equation}\label{105}
 \Phi^{\dot\a\a}\Rightarrow(T,\;Q)\oplus (M, \;\bar M).
\end{equation}

As to the three complex constraints $\Psi^{\a\b}$  they are already presented in the abelian form          
\begin{equation}\label{106}
\{\Psi^{\a\b},\;\Psi^{\g\d} \}_{P.B.}=0, \;\; \{ \Psi^{\a\b},\;{\bar \Psi}^{\dot\g\dot\d} \}_{P.B.}=0.
\end{equation}
and are  good candidates for the next set of the second class constraints. 
These $\Psi$-constraints may be unified with the two complex constraints \f{6} and \f{78}
\begin{equation}\label{107}
u^\a v_\a=1 ,\quad \,\; u^{\prime\a} u_\a =0.
\end{equation}
 to form a larger set including five complex second class constraints $Y^{A} \;(A=1,2,3,4,5)$  
\begin{eqnarray}\label{108}
\nonumber
Y^{A}\equiv(\Psi^{\a\b},\; u^\a v_\a-1, u^{\prime\a}u_\a)=0,\;\\
{\bar Y}^{\bar A}\equiv({\bar \Psi}^{\dot\a\dot\b},\; {\bar u}^{\dot\a}{\bar v}_{\dot\a}-1, {\bar u}^{\prime\dot\a}{\bar u}_{\dot\a})= 0.
\end{eqnarray}
which have zero P.B. between themselves 
\begin{equation}\label{109}
\{Y^{A},\; Y^{B}\;  \}_{P.B.}=0, \;\;  \{ Y^{A}, \; {\bar Y}^{\bar B} \}_{P.B.}=0.
\end{equation}
 
At last, going over to the constraints \f{35} ($P_{u}^{\a}=P_{v}^{\a}=0$) 
and observing that the projection $P_{v}^{\a}u_{\a}$ and its complex conjugate
 form the first class constraints $W, \; \bar W$ \f {53} one can choose  the complex projection
\begin{equation}\label{110}
P_{v}^{\a}v_{\a}=0
\end{equation}
together with the two complex constraints  
\begin{equation}\label{111}
P_{u}^{\a}=0
\end{equation}
as a new set of the three complex second class constraints $ \Xi^\L  \;(\L=1,2,3)$
\begin{eqnarray}\label{112}
\nonumber
 \Xi^\L\equiv( P_{u}^{\a},\; P_{v}^{\b}v_{\b})=0,\\
{\bar\Xi}^{\bar\L}\equiv( {\bar P}_{u}^{\dot\a},\; {\bar P}_{v}^{\dot\b}v_{\dot\b})=0.
\end{eqnarray}
The three complex constraints $\Xi^\L$ and ${\bar \Xi}^{\bar\L}$  have zero P.B. between themselves 
\begin{equation}\label{113}
\{ \Xi^{\L},\;  \Xi^{\S} \}_{P.B.}=0, \;\;  \{ \Xi^{\L}, \; {\bar\Xi}^{\bar\S}
\}_{P.B.}=0.
\end{equation}

We have thus, that the total set of second class constraints following from the action \f{22} may be divided into three complex abelian subsets:\\
the abelian ${\mathbf {M}}$-subset including one complex constraint
\begin{eqnarray}\label{114}
\nonumber
{\mathbf M}\equiv (M, \;\bar M),\\
\{{\mathbf M}, \;{\mathbf M}\}_{P.B.}=0;
\end{eqnarray}
the  abelian ${\mathbf {Y^A}}$-subset including five complex constraint
\begin{eqnarray}\label{115}
\nonumber 
{\mathbf {Y^A}}\equiv (Y^{A}, \;\bar Y^{\bar A}),\\
\{{\mathbf  {Y^A}}, \;{\mathbf {Y^B}}\}_{P.B.}=0;
\end{eqnarray}
the abelian ${\mathbf {\Xi^\L}}$-subset including three complex constraint
\begin{eqnarray}\label{116}
\nonumber 
{\mathbf {\Xi^\L}}\equiv (\Xi^{\L}, \;\bar\Xi^{\bar\L}),\\
\{ {\mathbf {\Xi^\L}}, \;{\mathbf {\Xi^\S}}\}_{P.B.}=0.
\end{eqnarray}
Having calculated the P.B. between these subsets
\begin{eqnarray}\label{117}
\nonumber
\{{\mathbf {Y^A}}, \;{\mathbf M}\}_{P.B.}= {\mathbf {G^A}}, \\
\nonumber
\{{\mathbf {Y^A}}, \;{\mathbf {\Xi^\L}}\}_{P.B.}={\mathbf F^{A\L}}, \\
\{ {\mathbf {\Xi^\L}}, \; {\mathbf M}\}_{P.B.}= {\mathbf {H^\L}}.
\end{eqnarray}
one can present the antisymmetric real 18x18 Dirac's $\hat{\mathbf C}-$matrix
 constructed of the P.B.  of the second class constraints in the compact form
 of an 9x9 antisymmetric complex matrix
\begin{eqnarray}\label{118}
\hat{\mathbf C}=\left( \begin{array}{lrc}\;\;\;{\mathbf 0}& {\mathbf F^{A\L}} \;\;\;& {\mathbf {G^A}} \\
-{\mathbf F^{A\L}}^{T}& {\mathbf 0}\;\;\;\;\;\;\;& {\mathbf {H^\L}}\\
-{\mathbf {G^A}}^{T}& -{\mathbf {H^\L}}^{T}\;\;\;& {\mathbf 0}\; \end{array}\right)
\end{eqnarray}

This  $\hat{\mathbf C}$-matrix is then used to construct the Dirac bracket
\begin{equation}\label{119}
\{f(\s),\;\; g(\s^\prime)\}_{D.B.}=\{f(\s),\; g(\s^\prime)\}_{P.B.} -
\sum\int
\{f(\s),\;\bullet\}_{P.B.}({\mathbf \hat C^{-1}})^{\bullet\bullet}\{\bullet \,,\; g(\s^\prime)\}_{P.B.}
\end{equation}
and to derive the correct Hamiltonian equation of motion defined by this  Dirac bracket
\begin{equation}\label{120}
\dot f(\t,\s) =\int d{\s^{\prime}}\{{\cal H}(\s^{\prime}),\;  f(\t,\s)\}_{D.B.}
\end{equation}
and the D.B. algebra resulting  from  the P.B. algebra \f{96}.

\section {Conclusion}

Here we constructed the Hamiltonian and studied the constraint structure in 
a new model of strings in a  four dimensional space-time extended by the addition of six real  TCC coordinates $z_{mn}$. A  covariant separation  of these constraints into  first and second classes has been described. We found that the 20 real primary  \f{6}, \f{31}, \f{33}, \f{35} and $2 $ secondary  \f{78} constraints are covariantly splitted into 4 real first  class constraints \f{92} and  18 real second class constraints \f{103}, \f{108} and \f{112}
$$20|_{primary\; constr.} + 2|_{secondary\; constr.} = 4|_{first \;class \;constr.}+  18|_{second \;class \;constr.}$$
The total number of the phase variables \f{28} and \f{29} in the model is 
$36=(18+18)$. Recalling  that the 4 first class constraints kill 8  phase variables, we conclude that  the number of  independent physical phase variables is $10= 36-(4+4+18)$. These 10 phase space  variables correspond to 5 off shell physical degree of freedom.  So, we conclude that the  TCC coordinates $z_{mn}$  together with the dyads $u_{\a}$ and  $v_{\a}$ contribute
in fact only one real off shell physical coordinate into addition to the 4 real world cooordinates $x_{\a\dot\a}$. Accordingly, the described string  
 moves in an effective 5-dimensional space-time.

The problem is now how to construct this 5-dimensional space-time. 

With this in mind we constructed  the P.B. algebra for the first class constraints. It  has  a structure  similar to that of the contracted rotation algebra 
of (Anti) de Sitter space-time. We found that this P.B. algebra coincides with the correspondent algebra for tensionless strings  and that discovered an essential role for  the second class constraints which  codes the physical effects associated with the TCC coordinates.
 A formulation in terms of the Dirac bracket algebra  (or an  equivalent construction \cite{fs,bff,em})  is needed to derive the covariant string  dynamics in the effective 5-dimensional space-time. 

Having the Dirac bracket constructed we are able to study  the Dirac bracket algebra for  the first class constraints and to describe the effective 5-dimensional space-time associated with the the TCC coordinates $z_{mn}$. 
We expect this effective space-time to be $AdS_{5}$.
Now we are in a progress to realize this goal.   

 After completion of this work D. Polyakov informed us that the addition of 
 a  5-form vertex operator corresponding to a brane-like state into the NSR superstring action, effectively curves the $D=10$ space-time to that of $AdS_{5}$x$S_{5}$
\cite{pol}. This vertex operator contains the  world sheet NSR fermions and plays the role of cosmological-like term.

\section {Acknowledgements}

A.Z. thanks Fysikum at the Stockholm University for the kind hospitality and Dima Polyakov for the useful discussion. The work is partially supported by the Axel Wenner-Gren Foundation.  A.Z. is partially supported by Award CRDF-RP1-2108 and by Ukrainian SFFR project 02.07/276. U.L. is supported in part by NFR grant 5102-20005711 and by EU contract HPRN-C7-2000-0122.


\begin{thebibliography}{99}

\bibitem{dafre}
R. D'Auria and P. Fre, \NPB 201 1982 101.
\bibitem{holpro}
J. van Holten and A. van Proyen, \JPA 15 1982 3763.
\bibitem{ziz} 
P.A. Zizzi, \PLB 149 1984 333. 
\bibitem{hupol}
J. Huges and J. Polchinski, \NPB 278 1986 147.
\bibitem{azto}
J.A. de Azcarraga and  P.K. Townsend, \PRL 62 1989 2579.
\bibitem{dust}
M. Duff and K.S. Stelle, \PLB 253 1991 113.
\bibitem{wit} 
E. Witten, \NPB 443 1995 85.
\bibitem{pol1} 
J. Polchinski, \PRL 75 1995 184.
\bibitem{town2}
 P.K. Townsend,  M-theory from its superalgebra; hep-th/9712004.
\bibitem{curt}
T. Curtright,
\PRL 60 1988 393. 
\bibitem{gren}
M.B. Green, \PLB 223 1989 157.
\bibitem{hew}
 S. Hewson, \NPB 507 1997 445; hep-th/9701011.
\bibitem{eis}
Y. Eisenberg and S. Solomon, \PLB 220 1988 562.
\bibitem{sieg}
W. Siegel, \PRD 50 1994 2799.
\bibitem{bes}
 E. Bergshoeff and E.Sezgin, \PLB 354 1995 256.
\bibitem{se}
 E. Sezgin, \PLB 392 1997 323.
\bibitem{azgit}
J.A. de Azcarraga, J.P. Gauntlett, J.M. Izquierdo and  P.K. Townsend, \\
\PRL 63 1989 2443.
\bibitem{ds} G. Dvali and M. Shifman, \PLB 396 1997  64; \\ erratum \PLB 407 1997 452; hep-th/9612128.
\bibitem{gs}
A. Gorsky and  M. Shifman,  More on the tensor central charges in  N=1 supersymmetric gauge theories ( BPS wall junctions and strings), hep-th/9909015.  
\bibitem{vy}
G. Veneziano and S. Yankielowicz,  \PLB 113 1982 231.
\bibitem{ruds}
I. Rudychev and E. Sezgin, \PLB 415 1997 363; hep-th/9711128.
\bibitem{balu}
I. Bandos and J. Lukierski, \MPLA 14 1999 1257.
\bibitem{ggt}
 P.K. Townsend and G.W. Gibbons,  \PRL 83 1999 1727.
\bibitem{gah}
 J.P. Gauntlett and C.M. Hull,  \JHEP 001  2000 004; hep-th/9909098.
\bibitem{gght}
 J.P. Gauntlett, G. Gibbons, C.M. Hull and P.K. Townsend, \\  \CMP  216 2001 431.
\bibitem{caib}
C. Chryssomalakos. J.A. de Azcarraga, J.M. Izquierdo, J.C. Perez Bueno,\\  \NPB 567 2000 293
\bibitem{zli} 
A.A. Zheltukhin and U. Lindstr\"om,  Strings in space with tensor central charge \\coordinates, \NPBP 102-103 2001 126.
\bibitem{gz} 
O.E. Gusev and A.A. Zheltukhin,  \JETP 64 1996 487.
\bibitem{ulf2}
 U. Lindstr\"om, B. Sundborg and G. Theodoridis, \PLB 235 1991 319.
\bibitem{lizzi}
F. Lizzi, B. Ray, G. Sparano and A. Srivastava, \PLB 182 1986 326.
\bibitem{guliss}
 H. Gustafsson,  U. Lindstr\"om, P. Salsidis, B. Sundborg  and R. von Unge,\\
\NPB 440 1995 495.
\bibitem{zi}
A.A. Zheltukhin,  \SJNP 51 1990 950;\\
 K. Ilienko and A.A. Zheltukhin,  \CQG 16 1999 383.
\bibitem{town1}
 P.K. Townsend,\PLB 277 1992 285.
\bibitem{belt} 
E.Bergshoeff, L.A. London and P.K. Townsend,  \CQG 9 1992 2545.
\bibitem{haliu} 
S. Hassani, U. Lindstr\"om and R. von Unge,  \CQG 11 1994 L79.
\bibitem{sch}
A. Schild,  \PRD 16 1977 1722.
\bibitem{kl} 
A. Karlhede and U. Lindstr\"om,  \CQG 3 1986 L73. 
\bibitem{z1}
 A.A. Zheltukhin,  \JETP 46 1987 672.
\bibitem{z2}
A.A. Zheltukhin, \SJNP 48 1988 950;\\
I.A. Bandos and A.A. Zheltukhin,  \FdP 41 1993 619.
\bibitem{lir}
U. Lindstr\"om and R. von Unge,
 \PLB 403 1997 233; hep-th/9704051;\\
H. Gustafson and U. Lindstr\"om,  \PLB 440 1998 43; hep-th/9807064;\\
U. Lindstr\"om, M. Zabzine and A.A. Zheltukhin,
 \JHEP \\ 12 1999 016; hep-th/9910159.
\bibitem{fs}
L.D. Faddeev and S. Shatashvili,   \PLB 167 1986 225.
\bibitem{bff}
I.A. Batalin, E.S. Fradkin and T.E. Fradkina,  \NPB 279 1987 514.
\bibitem{em}
E.S. Egorian and R.P. Manvelian, Preprint YERPHI-1056(19)-88 (1988), \\ Yerevan 1988.
\bibitem{z5} 
A.A. Zheltukhin, \TMP 77 1988 377.
\bibitem{pol}
D. Polyakov,  $D=4$ Gauge theory correlators  from $D=10$ NSR $\s-$model,\\
 hep-th/9907021;\\
D. Polyakov,  Ads/CFT Correspondence, critical strings and stochastic quantization,
 hep-th/0005094.
\end{thebibliography}
\end{document}